\documentclass[5p,sort&compress,times,preprint]{elsarticle}
\journal{Computer Physics Communications}
\pdfoutput=1
\usepackage{graphicx}
\usepackage{amsmath}

\usepackage[frozencache]{minted}
\newminted{cpp}{numbers=left,numbersep=8pt,fontsize=\footnotesize,xleftmargin=0.05\columnwidth,xrightmargin=0.04\columnwidth}
\newminted{bash}{numbers=left,numbersep=8pt,fontsize=\footnotesize,xleftmargin=0.05\columnwidth,xrightmargin=0.04\columnwidth,breaklines}
\newminted{json}{numbers=left,numbersep=8pt,fontsize=\footnotesize,xleftmargin=0.05\columnwidth,xrightmargin=0.04\columnwidth}
\newminted{mathematica}{numbers=left,numbersep=8pt,fontsize=\footnotesize,xleftmargin=0.05\columnwidth,xrightmargin=0.04\columnwidth,breaklines}
\newmintinline[bash]{bash}{fontsize=\small}
\newmintinline[cpp]{C++}{}
\newmintinline[json]{C++}{}
\newmintinline[mathematica]{mathematica}{}
\newcommand{\path}[1]{\cpp{#1}}

\usepackage[pdfborder={0 0 0}]{hyperref}
\usepackage[utf8]{inputenc}

\renewcommand{\vec}[1]{\mathbf{#1}}
\newcommand{\Tr}{\operatorname*{Tr\,}}
\renewcommand{\th}{\textsuperscript{th} }

\begin{document}

\begin{frontmatter}
\title{High-Temperature Series Expansion for Spin-1/2 Heisenberg Models}

\author[eth]{Andreas Hehn}
\ead{hehn@phys.ethz.ch}

\author[psi]{Natalija van Well}
\ead{natalija.van-well@psi.ch}

\author[eth]{Matthias Troyer}
\ead{troyer@phys.ethz.ch}

\address[eth]{Theoretical Physics, ETH Zurich, CH-8093 Zurich, Switzerland}
\address[psi]{Laboratory for Neutron Scattering and Imaging, Paul Scherrer Institute, CH-5232 Villigen, Switzerland}

\date{\today}

\begin{abstract}
We present a high-temperature series expansion code for spin-1/2
Heisenberg models on arbitrary lattices.
As an example we demonstrate how to use the application for an anisotropic
triangular lattice with two independent couplings $J_1$ and $J_2$ and calculate
the high-temperature series of the magnetic susceptibility and the static
structure factor up to 12\th and 10\th order, respectively.
We show how to extract effective coupling constants for the triangular
Heisenberg model from experimental data on Cs$_2$CuBr$_4$.
\end{abstract}

\begin{keyword}
%% keywords here, in the form: keyword \sep keyword
series expansions; quantum magnetism; triangular Heisenberg model
\end{keyword}

\end{frontmatter}
{\bf PROGRAM SUMMARY}
  %Delete as appropriate.

\begin{small}
\noindent
%{\em Manuscript Title:} High-Temperature Series Expansion for Spin-1/2
%     Heisenberg Models \\
{\em Program Title:} LCSE \\
{\em Authors:} Andreas Hehn, Matthias Troyer\\
{\em Licensing provisions:} Use of the applications or any use of the source
code requires citation of this paper.\\
{\em Programming language:} \verb*#C++11#, MPI for parallelization, Mathematica
for analysis of results. \\
{\em Computer:} PC, cluster, or supercomputer \\
{\em Operating system:} Any, tested on Linux\\ 
{\em RAM:} 1 GB - 100 GB.\\ 
{\em Number of processors used:} 1 - 4096.\\ 
{\em Keywords:} linked cluster expansion, high-temperature series, series
expansions, Heisenberg model.\\
  % Please give some freely chosen keywords that we can use in a
    % cumulative keyword index.
{\em Classification:} 7.7  \\ 
  %Classify using CPCll embeddable graphs with n ≤ N Program Library Subject Index, see (
            % http://cpc.cs.qub.ac.uk/subjectIndex/SUBJECT_index.html)
    %e.g. 4.4 Feynman diagrams, 5 Computer Algebra. 7.7 Other Condensed Matter
    %inc. Simulation of Liquids and Solids
{\em External routines/libraries:}  ALPS \cite{ALPS1,ALPS2,ALPSWEB}, GMP \cite{GMP}\\
{\em Nature of problem:}
Calculation of thermodynamic properties (magnetic susceptibility and static
structure factor) for quantum magnets on arbitrary lattices.
A particularly hard problem pose quantum magnets on so frustrated lattice
geometries, as they can not be solved efficently by Quantum Monte Carlo methods.
\\
{\em Solution method:}
High-temperature series expansions employing a linked-cluster expansion allow to
obtain a high-order series in the inverse temperature for thermodynamic
quantities in the thermodynamic limit.
The resulting high-temperature series are exact up to the expansion order.
We implement the calculation of high-temperature series for the zero-field
magnetic susceptibility and static magnetic structure factor for the spin-1/2
Heisenberg model on arbitrary infinite lattices in arbitrary dimension.
\\
{\em Program code and examples:}
\url{http://www.comp-phys.org/lcse/} \\
\end{small}

% ============================================================================
\section{Introduction}
% ============================================================================

%Low dimensional quantum antiferromagnets are a major topic in condensed
Quantum antiferromagnets in low dimensions are a major topic in condensed
matter physics. The initial reason for intensive study of antiferromagnetic
systems was the discovery of antiferromagnetic order in the copper oxide layers
of undoped parent compounds of high-temperature superconducting
cuprates~\cite{vakin}. Since then, research on low dimensional antiferromagnetic
structures has evolved into an independent field because these systems are
strongly affected by quantum fluctuations and offer a great variety of exotic
phases, like valence bond solids or quantum spin liquids~\cite{valencebondsolid}.
From a theoretical perspective they are described by Heisenberg models,
which are, due to their simplicity and the many exotic phases they exhibit,
one of the most important class of toy models to study quantum phase
transitions~\cite{sachdev}. Furthermore, they can also serve a well controllable
environment to investigate more general phenomena like Bose-Einstein
condensation~\cite{giamarchi}.
There exist plenty of experimental realizations for quantum magnets, such as the
undoped La$_2$CuO$_4$, NaTiO$_2$ \cite{NaTiO2} or superconducting organic molecular
crystals \cite{organic}, to name a few.
A common problem is the connection of the experiments to the theoretical models,
i.e.\ the determination of coupling constants of theoretical models for a given
material~\cite{Coldea1,esr}.

An easy link between experiment and theory can be established by high-order
high-temperature series expansions for the microscopic models.
High-temperature series expansions have a long tradition in
condensed matter physics~\cite{DombGreen,oitmaabook}, and complement other
numerical methods, such as exact diagonalization or quantum Monte Carlo,
which are limited to small system sizes or non-frustrated systems, respectively.
For translational invariant lattices, high-temperature series expansions provide
 results directly for the thermodynamic limit.
The method is applicable to both simple bipartite spin models, as well as
geometrically frustrated models or fermionic models~\cite{pryadko} in arbitrary
dimensions.
The only approximation of the method is a finite order of the series.

In this paper we present a collection of applications to compute
high-temperature series expansions for Heisenberg models on arbitrary lattices.
As an example we compute the high-temperature series for an Heisenberg model on
a triangular lattice with spatial anisotropy and show how to obtain estimates
for the effective coupling constants for Cs$_2$CuBr$_4$.
Previous studies of this model by Zheng et al.\ \cite{Zheng} derived
the 10\th order series for the uniform magnetic susceptibility.
We add two additional orders to this series and derived 10\th order
series for the static magnetic structure factor.

% ============================================================================
\section{Model}
\begin{figure}
    \begin{center}
    \includegraphics[width=0.5\columnwidth]{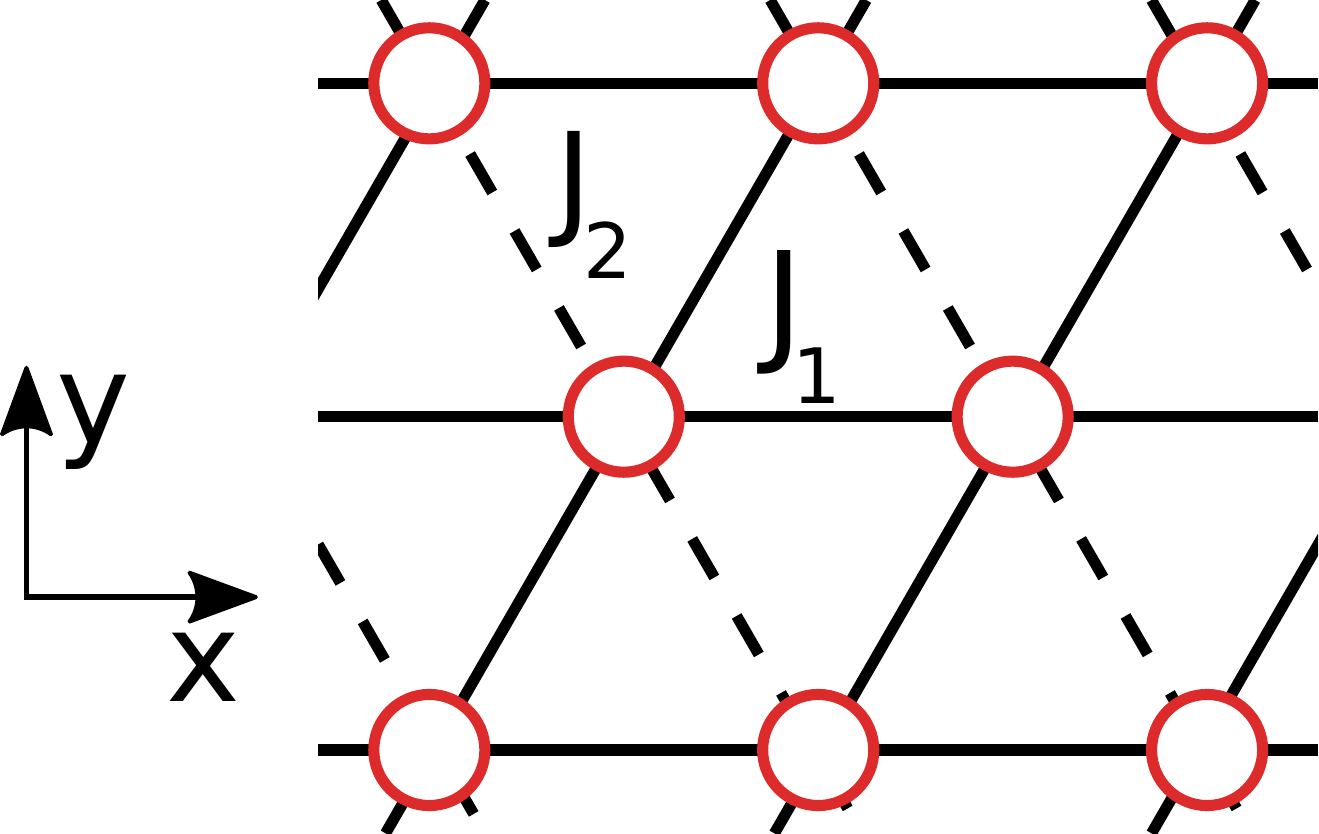}
    \caption{Triangular lattice with spatial anisotropy: exchange coupling
        $J_1$ along the solid lines and $J_2$ along the dashed lines.}
    \label{fig:lattice}
    \end{center}
\end{figure}

% ============================================================================
The spin-$1/2$ Heisenberg model is a lattice model where each site
is occupied by a spin $i$ spanning a local Hilbert space
$\mathcal{H}_i = \{|\uparrow \rangle , | \downarrow \rangle\}$.
The spins couple along bonds of the lattice and are governed by the Hamiltonian
\begin{equation} \label{eq:hamiltonian}
    H = \sum_{\langle i,j \rangle} J_{i,j} \, \vec S_i \vec S_j \, ,
\end{equation}
where $\vec S_i$ are the spin-$1/2$ operators acting on the spin $i$,
and $J_{i,j}$ are coupling constants for each bond $\langle i, j \rangle$ of the
lattice.
For the models considered, we only consider a small set of independent
coupling constants $J_{i,j} \in \{ J_1, J_2, \dots \}$.
Bonds with the same coupling constant are said to be of the same type.
While our application can handle up to four different bond types,
we here focus on a model with only two bond types, namely the Heisenberg model on
an anisotropic triangular lattice depicted in figure~\ref{fig:lattice}.
With two freely tunable constants $J_1,J_2$ the model covers three well known
special cases: for $J_1 = 0$ the model decomposes into uncoupled spin chains,
for $J_2 = 0$ the model is equivalent to the Heisenberg model on a square
lattice, and for $J_1 = J_2$ it becomes an isotropic triangular lattice.

The high-temperature series we obtain are symbolic polynomials in the variables
$\beta J_1$, $\beta J_2$. As the coupling constants remain freely tunable our
series is valid for both the ferromagnetic ($J_1,J_2 < 0$) and the
antiferromagnetic case ($J_1,J_2 > 0$).
In our analysis of the results we focus on the latter, as this case leads to
geometric frustration on the triangular lattice as not all bonds can be
satisfied classically.
This enhances quantum fluctuations and promises an interesting phase diagram.
Furthermore, the geometric frustration of the model makes it hard to study the
system with the otherwise successful method of Quantum Monte Carlo (QMC) due
to the sign-problem caused by the frustration.
Hence the model renders an ideal playground for our high-temperature series
expansions.

% ============================================================================
\section{High-temperature Series Expansions}
% ============================================================================

The high-temperature series expansions (HTE) of a
thermodynamic quantity $Q$ in the inverse temperature~$\beta$ can be written as
\begin{equation}
    Q(\beta) = \sum_n a_n \beta^n \,.
\end{equation}
For the Heisenberg model the quantities of interest are in many cases the
partition function
\begin{equation}
    Z = \Tr_{\mathcal{H}} \mathrm{e}^{-\beta H}
    \approx \sum_{n=0}^N \frac{(-\beta)^n}{n!}
    \Tr_{\mathcal{H}} H^n \, ,
\end{equation}
the uniform magnetic susceptibility
\begin{equation}
    \chi = \beta^{-1} \frac{\partial^2}{\partial h^2} \ln Z \,,
\end{equation}
and the spin-spin correlators
\begin{equation}
    \langle S_i^z S_j^z \rangle = \frac{1}{Z}
      \Tr_{\mathcal{H}} S_i^z S_j^z
         \mathrm e^{-\beta H}\,.
\end{equation}
which are Fourier transformed to obtain the static structure factor
\begin{equation}
   S(\vec k) = \frac{1}{N} \sum_{i,j} \langle S_i^z S_j^z \rangle e^{-i \vec k
       (\vec x_j - \vec x_i)} \, .
\end{equation}
The trace in those equations runs over the full Hilbert space $\mathcal{H} =
\mathcal{H}_1 \otimes \mathcal{H}_2 \otimes \dots$ of
the quantum system.
In the thermodynamic limit, this Hilbert space is of infinite dimension, as
the lattice is infinite.
However, for the series expanded term we may still obtain results in the
thermodynamic limit.
The limitations are a finite truncation order $N$ and a finite interaction
range of $H^n$-terms, when $H$ is composed of local interactions.

To calculate the series we use  the linked-cluster expansion, which expresses an
extensive quantity $A$ on the lattice $L$ as a sum of so-called irreducible
``weights'' $w^{(g)}$, of small connected parts of the lattice, the embeddable
connected graphs or clusters $g \in G_L$.
As a graph $g$ can appear many times in the lattice the weight $w^{(g)}$ has to
be multiplied by number of distinguishable embeddings in the lattice, the
lattice constant $\mathrm{LC}(g)$
\begin{equation} \label{eq:lattice_sum}
    A = \sum_{g \in G_L} \mathrm{LC}(g) \cdot w^{(g)} \, .
\end{equation}
For infinite, translational invariant lattices the quantity $A$ is usually
expressed per site, in which case the embeddings are also counted per site.

The same idea applies to the graphs themselves.
The extensive quantity $A^{(g)}$ defined on a graph $g$ is again expressed
as a sum of contributions over all embeddable graphs $g'$
\begin{equation} \label{eq:graph_sum}
    A^{(g)} = \sum_{g' \subset g} c(g',g) \cdot w^{(g)} \, .
\end{equation}
Similar to the lattice constant, $c(g',g)$ denotes the number of ways that the
graph $g'$ can be embedded in the graph $g$. Note that this sum includes the
graph $g$ itself.
Starting from a graph $g_1$ of only a single vertex without any edges, where
\begin{equation}
    A^{(g_1)} = w^{(g_1)} \, ,
\end{equation}
equation \ref{eq:graph_sum} defines a recursive scheme to determine the
irreducible weight $w^{(g)}$ in the so-called subcluster-subtraction:
given an extensive quantity $A^{(g)}$ of each graph $g$ we can obtain the
irreducible weights $w^{(g)}$ by recursively subtracting the weights $w^{(g')}$
of all graphs $g'$ embeddable in $g$.
The weights $w^{(g)}$ may then be combined to $A$ defined on the full lattice
in equation (\ref{eq:lattice_sum}).

For an infinite lattice the number of contributing graphs $g$ is evidently
infinite.
However, due to the finite range of the $H^n$-terms in a high-temperature series
expansion all irreducible weights $w^{(g)}$ for graphs with more than $n$ edges
hosting a pair-wise interaction, such as the $\vec S_i \vec S_j$ interaction of
the Heisenberg model, vanish.
Hence, for a finite expansion order $N$ we only need to consider a small
fraction of all embeddable graphs $G_L$, namely all embeddable graphs with
$n \le N$ edges.

Putting the pieces together, the computation of an high-temperature series
consists of four stages:
\begin{enumerate}
\item finding all contributing connected graphs
\item computing the extensive quantity $A^{(g)}$ for each graph
\item performing a subcluster-subtraction to obtain irreducible
weights $w^{(g)}$
\item embedding the graphs in the lattice and summing up the weights
accordingly.
\end{enumerate}

% ============================================================================
\section{Running the Application}
% ============================================================================

Similar to those four stages running our code for the spin-1/2 Heisenberg model
on a given lattice involves
three steps, each carried out by a dedicated application as shown in the
    respective example calls:
\begin{enumerate}
\item the graph generation, creating a list of all graphs embeddable into the
    desired lattice
\begin{bashcode}
graph_generator latticeparam.json graphlist.graphs
\end{bashcode}
\item the weight computation, calculating the contribution to the free energy
    series, magnetic susceptibility series and spin structure factor series for
    each graph
\begin{bashcode}
spin1_2_heisenberg_hte_compute_weights_2_variables 10 graphlist.graphs raw_heisenberg.series
\end{bashcode}
\item and finally the embedding step, performing the subcluster-subtraction and
    embedding into the lattice.
\begin{bashcode}
spin1_2_heisenberg_hte_reduce_embed_weights latticeparam.json graphlist.graphs raw_heisenberg.series reduced_heisenberg_series
\end{bashcode}
\end{enumerate}
In the following sections we explain these steps and the associated applications
in more detail.

\subsection{Graph Generation}
The first program
\begin{bashcode}
graph_generator <parameterfile> <outputfile>
\end{bashcode}
generates all embeddable graphs with up to $N$ edges for a lattice specified
using the ALPS lattice library~\cite{ALPS1,ALPS2}.
Starting from a single vertex it generates graphs by recursively adding edges,
checking for already seen isomorphic graphs using a variant of the McKay
algorithm~\cite{McKay,Hartke} and trying to embed these graphs in the lattice.
The program takes two arguments: 
\begin{itemize}
\item \bash{<parameterfile>}, the name of a parameter
file in JSON format, and 
\item \bash{<outputfile>}, the name of the output file in
which the found graphs are stored.
\end{itemize}

Listing~\ref{lst:parameterfile} shows the parameter file used for the
anisotropic triangular lattice.

\begin{listing}[t]
\begin{jsoncode}
{
  "lattice": {
    "lattice_library" : "custom_lattices.xml",
    "type": "anisotropic triangular lattice 2 couplings",
    "L": 40,
  },
  "num_edges": 12,
  "homogeneous_model" : true
}
\end{jsoncode}
\caption{The parameter file for the \cpp{graph_generator} program used in our example.}
\label{lst:parameterfile}
\end{listing}

The first parameter, \json{"lattice"}, describes the lattice for which the
embeddable graphs are generated. The lattices are provided by the ALPS lattice
library which offers a big selection of predefined lattice and is easily
extensible.

The parameters \json{"type"} and \json{"L"} are the name and the extent of the
lattice and are passed through to the ALPS lattice library.
As our example relies on a lattice which is not part of the standard ALPS
lattice library, a custom library file is loaded via the
\json{"lattice_library"} parameter. Note that \json{"L"} is not the size of the
(infinite) lattice used in the calculations, but indicates a finite lattice
large enough so that embedded clusters starting at the center do not extend to
the boundary.

The next parameter, \json{"num_edges"}, sets the upper limit on the number of
edges of the generated graphs and should match the desired order of the series
to be computed.

The last parameter tells the graph generator that all terms of the
Hamiltonian, regardless of the edge type, are of the same form
($\vec S_i \vec S_j$).

For the anisotropic triangular lattice depicted in figure~\ref{fig:lattice}
we found $4'821'837$ embeddable graphs with $n \le 12$ edges.

\subsection{Weight Computation}
The weight computation is the most expensive part of the procedure.
Due to the rapidly growing number of graphs for higher orders, as well as the
exponentially growing Hilbert space with the number of sites in a graph, the
computational effort grows exponentially with the desired order of the series.
We provide applications for up to four independent coupling constants
\begin{bashcode}
spin1_2_heisenberg_hte_compute_weights_1_variables
spin1_2_heisenberg_hte_compute_weights_2_variables
spin1_2_heisenberg_hte_compute_weights_3_variables
spin1_2_heisenberg_hte_compute_weights_4_variables
\end{bashcode}
Each application accepts the following three parameters
\begin{bashcode}
... [--suscept-only] <order> <graphfile> <outputfile>
\end{bashcode}
where the first mandatory parameter \bash{<order>} is the desired order of the
expansion, followed by \bash{<graphfile>}, the name of the file
containing the list of graphs for which the weights should be computed, and
\bash{<outputfile>}, the name of the output file.
For each graph the application will compute the series for the partition
function, the susceptibility and the equal-time $S^z_i S^z_j$ correlator for the
static structure factor $S(\vec k)$. All results will be appended to the
combined output file.
If the structure factor is not of interest,
the mandatory arguments can be preceded by the optional switch
\bash{--suscept-only}, telling the application to
omit the calculation of the correlator series. This
allows to focus on pushing the susceptibility to the highest possible order.

If the output file \bash{<combinedseriesdb>} already exists, the program will
only compute and append the weights of the missing graphs. Therefore the
computations can be interrupted at any time and may be continued by re-executing
the same command.

For short series the regular applications for desktop computers
suffice. However, two to three additional orders may be achieved running the
MPI-parallelized versions of the applications on a cluster or supercomputer.
The MPI applications will group the graph contributions to be computed
in tasks of a few graphs each and schedule those tasks across all available MPI
processes.
The applications, which have to be executed within an MPI environment, share the
same command line arguments as their non-MPI counterparts,
but take three additional optional parameters controlling the
parallelization:
\begin{bashcode}
... [--suscept-only] <order> <graphfile> <outputfile> [threadspp] [checkp_interval] [task_size]
\end{bashcode}
\begin{itemize}
\item \bash{threadspp} is the number of threads per MPI process (default: 16).
    In most cases a single MPI process per cluster node should be used, running
    one thread per CPU-core of the node.
\item \bash{checkp_interval} sets the interval of checkpoints in seconds
(default: 30). At the end of each interval the completed computation tasks will
be collected from all MPI processes and stored in the output database.
\item \bash{task_size} controls the size of a computation task, i.e.\ the number
of graphs per computation task (default: 15).
Note that the program will only save completed tasks, where all graph
contributions have been calculated, to the database at the checkpoints.
Large \bash{task_size} and \bash{checkp_interval} values may therefore result in
dropping many computed contributions in incomplete tasks at termination of the
program. Very small values, on the other hand, may cause considerable network
traffic on the cluster.
\end{itemize}
If these optional arguments are not specified their default value will be used.

\subsection{Subcluster-Subtraction and Embedding}

Once the raw weights are calculated for all contributing graphs, the
\begin{bashcode}
spin1_2_heisenberg_hte_reduce_embed_weights <parameterfile> <graphfile> <combinedseriesdb> <seriesdb_prefix>
\end{bashcode}
program will perform the
subcluster-subtraction and sum up the irreducible weights according to the
embeddings of the graphs in the the lattice.
For this task the program requires four arguments:
\begin{itemize}
\item \bash{<parameterfile>}, the name of the
json file containing the lattice parameters,
\item \bash{<graphfile>} the name of the
file containing the list of contributing graphs for the lattice,
\item \bash{<combinedseriesdb>}, the name of the file with the combined raw weights for
the partition function, the susceptibility and optionally the two-site
correlator for all contributing graphs from the previous step, and
\item \bash{<seriesdb_prefix>}, a prefix for the intermediate files the
program will generate.
\end{itemize}
From the combined raw weights file the program
generates three files containing the full graph contributions before the
subcluster-subtraction:\\
\begin{tabular}{lp{3cm}}
 \bash{<sp>_raw_mBetaF.series} & free energy ($-\beta F$)\\
 \bash{<sp>_raw_suscept.series} & susceptibility\\
 \bash{<sp>_raw_SzSz_correlator.series} & $S^z_i S^z_j$ correlator
\end{tabular}\\
and three files with the irreducible weights after the subcluster-subtraction:\\
\begin{tabular}{lp{3cm}}
 \bash{<sp>_red_mBetaF.series} & free energy ($-\beta F$) \\
 \bash{<sp>_red_suscept.series} & susceptibility\\
 \bash{<sp>_red_SzSz_correlator.series} & $S^z_i S^z_j$ correlator.
\end{tabular}\\
The prefix \bash{<sp>} of the filenames is substituted by the value of the
command line argument \bash{<seriesdb_prefix>} mentioned above.
 
Once more, if the output file already exists any already calculated quantity
will be taken into account and only missing weights will be appended. As the
weights of the graphs are independent of the underlying lattice, the
weight files may contain weights for graphs not contributing to the lattice.
Once the subcluster-subtraction is completed and all irreducible weights are
available, the program tries to embed all graphs from the graph list file into
the lattice.
When the embedding is completed it will print out the resulting series in
Mathematica compatible format to standard-out.

% ============================================================================
\section{Series Analysis}
% ============================================================================
Analysis of the high-$T$ series is done in Mathematica, which allows us to
easily perform symbolic manipulations and has many of the needed tools, most
notably Pad\'e approximants, built-in.
Along with the high-temperature series expansion application code we ship two
Mathematica packages\footnote{padeanalysis.m and paderesidualanalysis.m\\
 found in /opt/lcse/share/lcse/data\_analysis after installation.}
which complement the built-in tools and provide
functions to simplify common tasks of the analysis.
Here we show how to extract estimates for the coupling constants for the
Heisenberg model from susceptibility measurements on Cs$_2$CuBr$_4$, using
Pad\'e approximants constructed from the high-temperature series.

\subsection{Pad\'e Approximants}

In most cases the bare series have a very limited convergence radius, which is
hard to estimate if only few coefficients are available.
A common method to extend the convergence radius of such a power series and to
provide a rough error estimate are Pad\'e approximants~\cite{Guttmann}.
Pad\'e approximants are a simple analytical continuation of the power series,
which, in contrast to the bare series, can represent simple poles in the
complex plane and therefore may better approximate the original function.
A Pad\'e approximant to a function $f(x)$ is a rational function
\begin{equation}
P[L,M](x) = \frac{a_0 + a_1 x + \dots + a_L x^L}{1 + b_1 x +
    \dots + b_M x^M}
\end{equation}
of numerator degree $L$ and denominator degree $M$, whose power series is
identical to the power series of the function $f(x)$ up to the order
$L+M$. The individual coefficients $a_i$, $b_j$ are uniquely defined by this
condition.
Error estimates from Pad\'e approximants have to be interpreted with care as
various approaches exist and none of them provides rigorous error
bars~\cite{GuttmannAndJensen,Glenister}.
A simple, yet effective method is to inspect the behavior and spread of the
Pad\'e approximants $P[L,N-L](x)$ for all $0 \le L \le N$ with increasing
order~$N$.
Approximants are trusted as long as most approximants of a given order~$N$
agree well. Higher confidence is usually put on balanced approximants, where
the numerator and denominator are of roughly same degree.

Special attention has to be payed to the pole structure of the approximant.
Approximants $P[L,M]$ exhibiting a close root-pole pair in the complex plane are
considered defective, since the pair would virtually cancel and add no further
information compared to a Pad\'e approximant of lower order.

\subsection{Extraction of Coupling Constants}

The effective coupling constants of the Heisenberg model for
a real quantum magnet can be estimated by fitting the Pad\'e approximants of the
magnetic susceptibility $\chi(T)$ to experimental data.
As an example we illustrate how to determine the coupling constants of the
Heisenberg model on an anisotropic triangular lattice for Cs$_2$CuBr$_4$.
The full procedure can be found in the Mathematica
notebook\footnote{examples/data\_analysis/Cs2CuBr4\_fitting.nb}
in the example folder of the code package.

In the beginning of the notebook we load the Mathematica packages shipped 
with the code and paste the high-temperature series $\chi/\beta$ copied from the
output of the series expansion application
\begin{mathematicacode}
<< "padeanalysis.m"
<< "paderesidualanalysis.m"
chiOverBeta = 1/4 - 1/8*K + (*...*) - 1/4*B + (*...*)
\end{mathematicacode}
where the the symbolic variables \mathematica{B} and \mathematica{K} correspond
to $\beta J_1$ and $\beta J_2$, respectively.

For the fit to experimental data, three free parameters of the susceptibility
function $\chi(T)$ of our model need to be determined:
the Heisenberg couplings $J_1$, $J_2$ and the $g$-factor.
As the couplings $J_1$ and $J_2$ appear as non-linear parameters in the Pad\'e
approximants, we determine these parameters by a parameter scan on an
equidistant grid, calculating the sum of squared residuals to the experimental
data for each grid point.

The range for the grid is derived from the behavior of the system at very high
temperatures, where the Curie-Weiss law
\begin{equation}
    \chi(T) = \frac{C}{T_{CW}+T}
\end{equation}
applies.
Here $\chi(T)$ is the magnetic susceptibility per mole, $C$ is the Curie
constant $C = N_A g^2 \mu_0 \mu_b^2/4k_B$ and $T_{CW}$ is the Curie-Weiss
temperature. 
From a fit to the experimental data we obtain a first estimate for the
$g$-factor and the Curie-Weiss temperature which relates to the coupling
constants of our system through mean-field theory
\begin{equation} \label{eq:tcw}
    T_{CW} = J_1 + J_2 / 2 \, .
\end{equation}
Depending on the temperature range considered, the fit of the Curie-Weiss law
for Cs$_2$CuBr$_4$ yields $1.94 < g < 2.03$ and
$3\mathrm{K} < T_{CW} < 20\mathrm{K}$ (not shown).
To benefit from this first estimate the free parameter $J_2$ in our analysis is
substituted by $T_{CW}$ according to eq.~\ref{eq:tcw}.

For the parameter scan we define an equidistant grid for
$5\mathrm{K} \le T_{CW} \le 20\mathrm{K}$ and
$0.1\mathrm{K} \le J_1 \le 15\mathrm{K}$ with a grid spacing of $0.1\mathrm{K}$
for each parameter.
The Mathematica packages, loaded before, offer a set of tools for this scan,
which are controlled by a common parameter set.
Such a parameter set is created with the
\mathematica{CreatePadeResidualLandscapesParameterSet} function from
a list of Pad\'e parameters $[L,M]$ and the ranges forming the two-dimensional
grid:
\begin{mathematicacode}
TcwRange = Range[5, 20, 1/10];
JRange = Range[1/10, 150/10, 1/10];
PadeParameters = AllPadeParametersOfOrder[12];
parameterset = CreatePadeResidualLandscapesParameterSet[ PadeParameters, TcwRange, JRange ];
\end{mathematicacode}

As Pad\'e approximants are univariate, the tools require a function
to reduce the multivariate high-temperature
series to a univariate series for the parameters of a grid point.
\begin{mathematicacode}
seriesforparameters =
  Function[ {tcw, J1},
    Evaluate[Simplify[
      chiOverBeta /. K -> ((2*tcw/J1 - 2)*B)
    ] ]
  ];
\end{mathematicacode}
The defined function \mathematica{seriesforparameters} takes the two parameters
of a grid point $T_{CW}$, $J_1$ and returns a univariate series in
\mathematica{B}~$=\beta J_1$, where all occurrences of \mathematica{K} were
substituted by \mathematica{K}~$=J_2/J_1\cdot$\mathematica{B}.
Internally the tools use this function to generate Pad\'e approximants
$P[L,M]($\mathematica{B}$)$ for fixed $T_{CW}$.

To derive a function for the susceptibility $\chi(T)$ with the physical units
from a Pad\'e approximant another (higher-order) function has to be defined.
This function takes the approximant $P[L,M]($\mathematica{B}$)$, the parameters
of the grid point $T_{CW}$ and $J_1$, as well as the experimental
\mathematica{data} and returns a function $\chi(T)$, where the temperature $T$
remains the only undetermined variable.
\begin{mathematicacode}
modelfromapproximant =
  Function[ {approximant, tcw, J1, data},
    Function[ {T},
      Evaluate[
        With[
          { fkt = g2*A*4*1/T*Simplify[
                approximant /. B -> J1/T
              ]
          },
          fkt /. FindFit[data, fkt, {g2}, T]
        ]
      ]
    ]
  ];
\end{mathematicacode}
In our example the function \mathematica{modelfromapproximant} creates 
$\chi(T)$ from an approximant by substituting \mathematica{B} for fixed $J_1$,
multiplying by the constant $4$\mathematica{A}~$= 4C / g^2$ and $1/T$,
and fitting the remaining
linear $g^2$ factor to the experimental data using Mathematica's
\mathematica{FindFit} function.

Finally we construct Pad\'e approximants and a corresponding model function
$\chi(T)$ for each grid point $T_{CW},J_1$ of the parameter scan and compute for
each function the sum of the squared residuals
\begin{equation}
  R = \sum_i (\bar \chi_i - \chi(T_i))^2
\end{equation}
against the experimental data points $(T_i, \bar \chi_i)$.
We named the sums of squared residuals on the grid for a fixed Pad\'e parameter
pair $[L,M]$ a residual landscape.
For the involved task of computing those residual landscapes for multiple Pad\'e
parameter pairs, the supplied Mathematica package offers a convenient
function
\begin{mathematicacode}
ComputeResidualLandscapes[seriesForParametersF, symbol, filterPredicates, modelFromPadeApproximantF, data, parameterset]
\end{mathematicacode}
The most important parameter is the \mathematica{parameterset} we encountered
before, which defines the grid and contains the list of Pad\'e parameters
$[L,M]$ for which the residual landscapes are calculated.
The function returns a list, with an residual landscape --- a two-dimensional
array containing the sums of squared residuals for each grid point --- for each
Pad\'e parameter pair $[L,M]$ of the parameter set.
The \mathematica{seriesForParametersF} parameter is the user defined function to
generate a univariate series for a grid point and \mathematica{symbol} is the
symbol of the variable of this series.
These arguments will be used to build Pad\'e approximants.
The \mathematica{filterPredicates} parameter takes a list of predicates which
are applied to the resulting approximants. Approximants not fulfilling the
predicates are immediately dropped.\\
The next parameter, \mathematica{modelFromPadeApproximantF}, defines how to build
a model function $\chi(T)$ from the approximant.
It expects a function taking four arguments, the approximant, the two parameters
of a grid point, and the data for the fit.
The last parameter \mathematica{data} is the set of experimental data points.

For the example the function is called as follows
\begin{mathematicacode}
residuals12thorder = ComputeResidualLandscapes[ seriesforparameters, B, {HasNoCloseRootPolePair[0.1]}, modelfromapproximant, experimentaldata, parameterset ];
\end{mathematicacode}
where \mathematica{HasNoCloseRootPolePair[0.1]} is a predicate to filter all
defective Pad\'e approximants, i.e.\ approximants a root-pole pair in the
complex plane with a distance less than $0.1$.

Once the computation is completed, \mathematica{residuals12thorder} contains
twelve residual landscapes for the Pad\'e parameters $[L,12-L]$,
$0 \le L \le 12$.
From this list of landscapes the best fitting Pad\'e of each grid point can be
selected by 
\begin{mathematicacode}
minimalresiduals = SelectMinima[residuals12thorder, parameterset ];
\end{mathematicacode}
returning a list with two elements, where the first element is the minimum sums
of squared residual for each grid point, and the second element is a
two-dimensional array of Pad\'e parameter pairs $[L,M]$ exhibiting the minimum
of the corresponding point.

% ============================================================================
\section{Experiments on Cs$_2$CuBr$_4$}
% ============================================================================
The Cs$_2$CuBr$_4$ crystals with an orthorhombic (Pnma) structure were grown
from aqueous solution with the evaporation method. This implies that with
respect to salts CsBr and CuBr$_2$ the 2:1 stoichiometric mixture was used for
the growth. For the crystal growth process a small evaporation rate such as
$25 \mathrm{mg}/\mathrm{h}$ was used to reach an appropriate crystallization
rate. This leads to a very good crystal quality. The sample was crystallized
within 4 weeks at room temperature \cite{Kruger}.
The magnetic susceptibility measurement of the Cs$_2$CuBr$_4$ single crystal was
determined in the temperature range between
$2\mathrm{K} \le T \le 300\mathrm{K}$ and in a magnetic field $B = 1\mathrm{T}$
using a Quantum Design SQUID magnetometer. The measurement was performed on a
single crystal, which was oriented parallel to the crystallographic $b$ axis.
The temperature-independent diamagnetic core contribution has corrected the
magnetic contribution of the sample holder and the constituents according to
\cite{Kahn}. The determination of the magnetic background is important,
especially at high temperatures, where the magnetic signal of this compound is
small.

% ============================================================================
\section{Results}
% ============================================================================

For the Heisenberg model on the anisotropic triangular lattice depicted in
figure~\ref{fig:lattice}, we obtained the high-temperature series
for the magnetic susceptibility up to 12\th order and the series for the static
spin structure factor up to 10\th order. The coefficients of the plain
high-temperature series can be found in table~\ref{tab:susseries} and the
example folder of the program sources, respectively.
In the following we compare the computed series to results from Quantum Monte
Carlo simulations on the same lattice for the case of an antiferromagnetic
square sub-lattice with ferromagnetic couplings along the diagonal chains
$J_1>0, J_2<0$ and we compare the fully antiferromagnetic case $J_1>0, J_2>0$
with experimental results on Cs$_2$CuBr$_4$.

\subsection{Comparison to Quantum Monte Carlo}
\begin{figure}
  \begin{center}
  \includegraphics[width=\columnwidth]{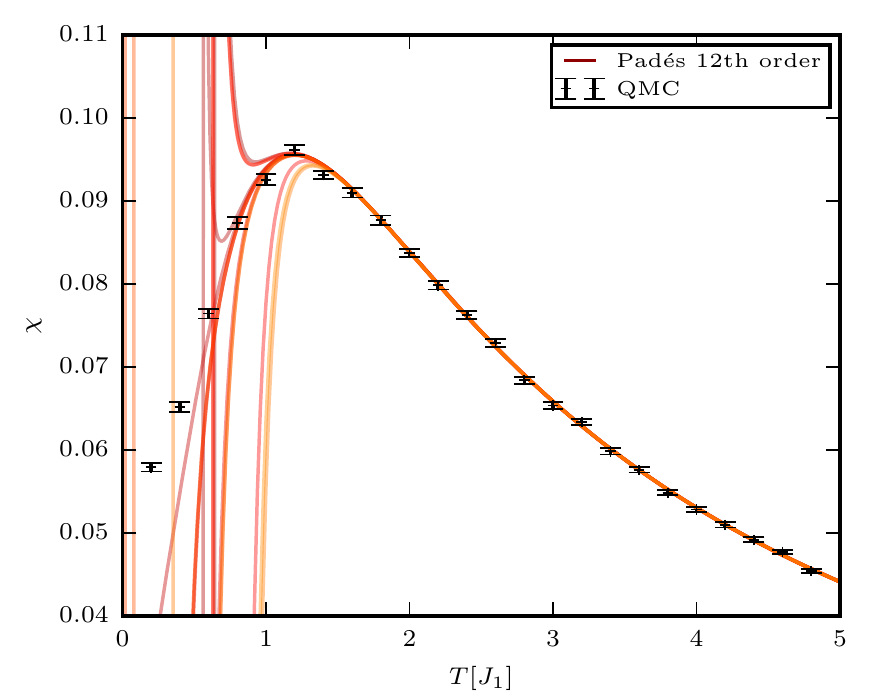}
  \caption{
      Comparison of Pad\'e approximants to the $12\th$ order series of the
      magnetic susceptibility $\chi(T)$ for a non-frustrated system with
      ferromagnetic chain couplings ($J_1 = 1, J_2 = -1$) with Quantum Monte
      Carlo simulations.
  }
  \label{fig:suscept_mc}
  \end{center}
\end{figure}
\begin{figure}
  \begin{center}
  \includegraphics[width=\columnwidth]{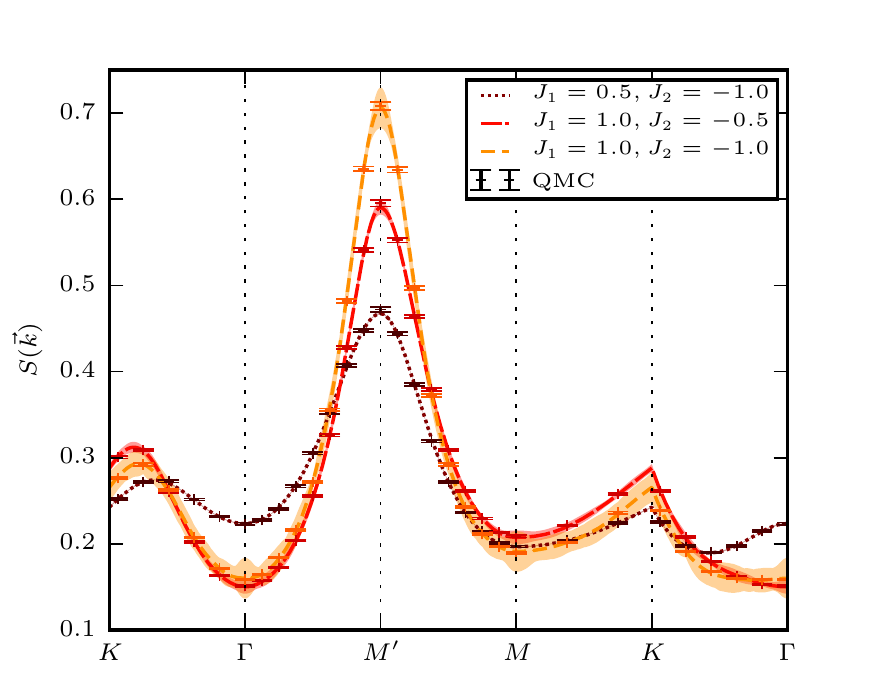}
  \caption{
      Comparison of the static structure factor $S(\vec k)$ computed by
      series expansions (lines) against Quantum Monte Carlo simulations (points
      with error-bars) for a non-frustrated systems with ferromagnetic chain
      couplings $J_1>0, J_2 < 0$ at $T=1.8\,J_1$.
      The path through the Brillouin zone is shown in figure~\ref{fig:structurefactor}.
      While discrete error-bars represent the standard deviation of Monte Carlo
      samples, the continuous error-areas around the lines represent the spread
      of Pad\'e approximants and are not rigorous error estimates.
  }
  \label{fig:structurefactor_mc}
  \end{center}
\end{figure}
In contrast to the fully antiferromagnetic case, the mixed case
with ferromagnetic couplings along the chains $J_1>0, J_2<0$ is not frustrated
and can be treated efficiently with Quantum Monte Carlo methods as it is not
subject to the well-known sign-problem.
We performed Quantum Monte Carlo simulations for three exemplary cases $J_1/J_2
= \{1/2, 1, 2\}$ using the \bash{dirloop_sse} application of
ALPS\cite{ALPS1,ALPS2,ALPSWEB}, which is based on the stochastic series expansion
(SSE) method.
Each data point is computed from $5\cdot 10^4$ samples on a lattice patch of
$64^2$ sites.
Figure~\ref{fig:suscept_mc} shows the temperature behavior of the magnetic
susceptibility for $J_1/J_2 = 1$ obtained from Quantum Monte Carlo together with
the non-defective Pad\'e approximants $P[L,M]$ with $11 \le L+M \le 12$
continuing the calculated 12\th order series.
Above $T \approx 1.2 J_1$ the spread of the Pad\'e approximants is negligible
and the approximants are in excellent agreement with the Monte Carlo results.
Below $T \approx J_1$ the different Pad\'e approximants start to spread
drastically.
Even in this temperature regime, down to $T\approx 0.6 J_1$, a few approximants
agree well with Monte Carlo.
In this regime, however, a prediction solely based on the computed series would
be hard to justify, due to the broad spread of the different approximants.
For the other parameters $J_1/J_2 = \{ 1/2, 2 \}$, a very similar spreading
behavior and agreement with Quantum Monte Carlo data of the Pad\'e approximants
is observed (not shown).

Also for the spin structure factor $S(\vec k)$
(fig.~\ref{fig:structurefactor_mc}), the 10\th order series and
Quantum Monte Carlo estimates agree beautifully down to $T\approx 1.8 J_1$ or rather
$T\approx 1.4 J_1$, depending on the particular couplings.
For lower temperatures the predictions from the series data become increasingly
unreliable as the Pad\'e approximants start spreading.
Both methods consistently display the peak at $M'$, which is a residue of
antiferromagnetic ordering at $T=0$.

\subsection{Comparison with Cs$_2$CuBr$_4$}
\begin{figure}
  \begin{center}
  \includegraphics[width=\columnwidth]{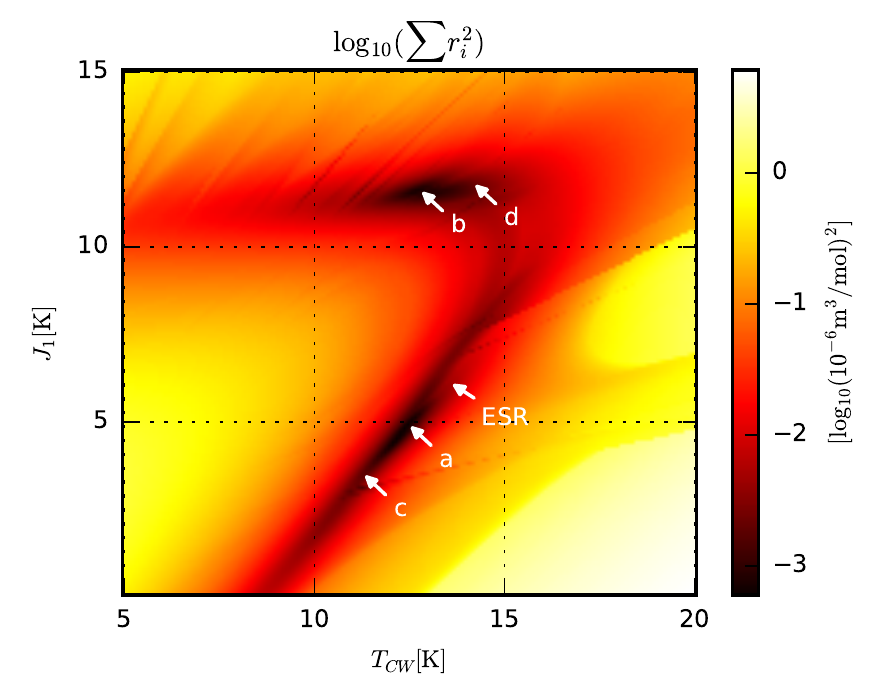}
  \caption{Sum of squared residuals $r^2$ of the best fitting Pad\'e approximants
      $P[L,M]$ for the computed magnetic susceptibility $\chi(T)$ compared to
      data for CS$_2$CuBr$_4$ with $T > 7K$. Minima:
      (a) $T_{CW}=12.5\mathrm{K}$, $J_1=4.9\mathrm{K}$,
      (b) $T_{CW}=12.8\mathrm{K}$, $J_1=11.6\mathrm{K}$;
      for comparison in figure~\ref{fig:pades}:
      (c) $T_{CW}=11.3\mathrm{K}$, $J_1=3.5\mathrm{K}$,
      (d) $T_{CW}=14.2\mathrm{K}$, $J_1=11.8\mathrm{K}$;
      (ESR) values obtained from ESR study $T_{CW}=13.6\mathrm{K}$,
          $J_1=6.1\mathrm{K}$~\cite{esr}.
  }
  \label{fig:residuals}
  \end{center}
\end{figure}

For Cs$_2$CuBr$_4$ we fitted the 12\th order series to the experimental data
on the magnetic susceptibility.
The model functions $\chi(T)$ with proper units are obtained from the bare
series
as described in the previous section.
Figure~\ref{fig:residuals} shows the sum of squared residuals compared to the
experimental data of the best fitting non-defective Pad\'e approximants $P[L,M]$
with $9 \le L+M \le 12$ for each point of the grid.
Note that we only consider $T > 7K$ for the procedure, as the approximants will
not approximate the underlying function well for low temperatures.
The cutoff temperature was chosen slightly below the point the different
approximants start to spread significantly.
Higher order calculations might reduce this spreading temperature further,
however, we do not expect significant changes with 1-2 additional orders, as
we see only marginal improvement from 10\th to 12\th order.
\begin{figure}
  \begin{center}
  \includegraphics[width=\columnwidth]{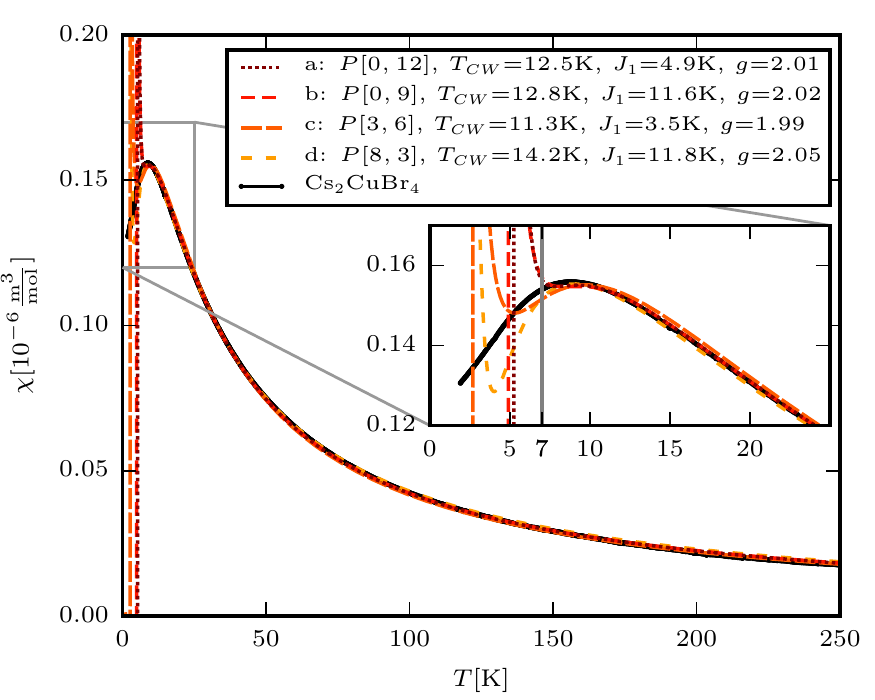}
  \caption{Pad\'e approximants fitted to Cs$_2$CuBr$_4$ data for $T > 7K$.}
  \label{fig:pades}
  \end{center}
\end{figure}

The parameter scan reveals two shallow elliptical valleys of the residuals
centered around the local minima at
$T_{CW} = 12.5\mathrm{K}$, $J_1 = 4.9\mathrm{K}$ and
$T_{CW} = 12.8\mathrm{K}$, $J_1 = 11.6\mathrm{K}$, respectively.
For both points the corresponding Pad\'es behave remarkably similar.
Both Pad\'es yield an excellent fit to the experimental data within the
considered temperature range (fig.~\ref{fig:pades}).
Also both Pad\'es slightly overestimate the susceptibility for high-temperatures
and slightly underestimate close to the maximum, before both Pad\'es diverge
around $T=6\mathrm{K}$.

However, also parameters at the verge of the valleys of the residual landscape,
e.g.\ at
$T_{CW}=11.3\mathrm{K}$, $J_1=3.5\mathrm{K}$ and
$T_{CW}=14.2\mathrm{K}$, $J_1=11.8\mathrm{K}$
( (c) and (d) in figures~\ref{fig:residuals} and \ref{fig:pades}),
fit the data very well and illustrate how a wide range of parameters along the
valleys leads to comparable fits.
The extracted $g$-factors of all presented approximants all lie within the range
expected from the Curie-Weiss law, but underestimate the $g$-factor obtained
from spin resonance experiments $g=2.09$~\cite{esr}.

\begin{figure}
  \begin{center}
  \includegraphics[width=\columnwidth]{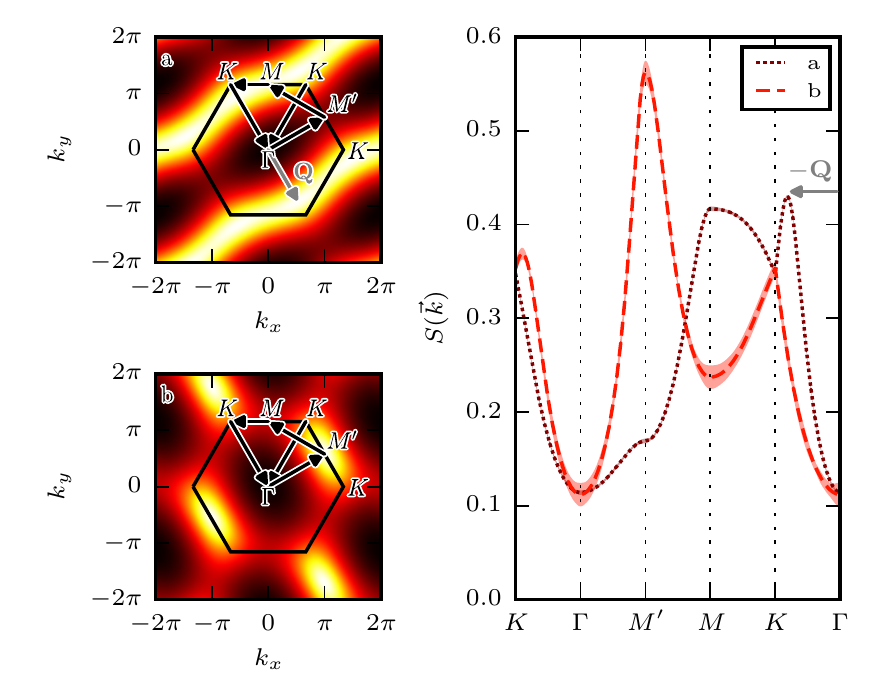}
  \caption{Static structure factor $S(\vec k)$ for
      (a) $T_{CW}=12.5\mathrm{K}$, $J_1 = 4.9\mathrm{K}$ and
      (b) $T_{CW}=12.8\mathrm{K}$, $J_1 = 11.6\mathrm{K}$
      at $T=15\mathrm{K}$. Error bars represent the spread of Pad\'e
      approximants and are not rigorous error estimates. For (a)
      $S(\vec k)$ exhibits a peak at $\vec Q=0.54 \vec b_{J_2}$
      hinting at incommensurate ordering.}
  \label{fig:structurefactor}
  \end{center}
\end{figure}

Physically the valleys correspond to two different regimes. In case (a)
the triangular model is dominated by the antiferromagnetic chain couplings
$J_2 = 15.2\mathrm{K}$ competing with a significantly smaller square lattice
coupling of only $J_1=4.9\mathrm{K}$ ($J_1/J_2 = 0.3(2)$).
This is also reflected in static structure factor $S(\vec k)$ shown in
figure~\ref{fig:structurefactor}, which features a plane wave structure along
the axis of the chain couplings, with maxima close to the
antiferromagnetic ordering vector of decoupled chains.
The plane wave is only slightly wiggled by the competing square lattice
couplings, pushing the maxima further to the corners of the Brillouin zone
to $\vec Q = 0.54 \vec b_{J_2}$, where $\vec b_{J_2}$ is the
reciprocal vector corresponding to the lattice vector $\vec a_{J_2}$ along the
$J_2$ bonds.
The other valley (b) corresponds to a triangular lattice with a strong square
lattice coupling $J_1 = 11.6\mathrm{K}$ perturbed by only a very small chain
coupling $J_2 = 2.4\mathrm{K}$ ($J_2/J_1 = 0.2(1)$).
Here the structure factor exhibits a peak at the symmetry point $M'$ as one
would expect from an antiferromagnetic ordering on the square lattice.

Previous studies of Cs$_2$CuBr$_4$ using high-temperature series expansions by
Zheng et al.~\cite{Zheng} obtained $J_1 = 6.99\mathrm{K}$ and
$J_2 = 14\mathrm{K}$ ($T_{CW} = 14\mathrm{K}$, $J_1/J_2 = 0.5$) from the 10\th
order series of the susceptibility, but also report that a wide range of ratios
$0.35 < J_1/J_2 < 0.55$ give comparable fits.
Those values reside at the upper end of our residual valley (a).
Another study extracted the Heisenberg couplings from high-field
electron-spin-resonance measurements with harmonic spin-wave theory~\cite{esr}.
Their findings of $J_1 = 6.1(3) \mathrm{K}$ and $J_2 = 14.9(7) \mathrm{K}$
corresponds to a Curie-Weiss temperature of $T_{CW} = 13.6(2) \mathrm{K}$ and
$J_1/J_2 = 0.41(0)$.
While also their coupling ratio seems significantly higher than the
$J_1/J_2=0.3(2)$ obtained here, the parameters lie well within valley (a)
(fig.~\ref{fig:residuals}) and are plausible candidates.

Further evidence for case (a) are neutron-scattering experiments on
Cs$_2$CuBr$_4$ \cite{Ono1,Ono2}, which revealed an incommensurate ordering
vector of $\vec Q = 0.575 \vec b_{J_2}$ at zero magnetic field, consistent with
our predictions from the high-temperature series for point (a)
(fig.~\ref{fig:structurefactor}).
While the predicted ordering vector for fit (a) $\vec Q = 0.54 \vec b_{J_2}$ is
slightly smaller than the ordering vector measured by neutron-scattering, one
has to bear in mind that they were obtained at significantly different
temperatures of $T=15\mathrm{K}$ and $T=60\mathrm{mK}$, respectively, and
might further converge.
Zheng et al.~\cite{Zheng} and Ono et al.~\cite{Ono2} also estimate the coupling
ratio $J_1/J_2$ from the ordering vector by comparison with various theoretical
models:
for the ground-state of the classical spin model on an anisotropic triangular
lattice, where neighboring spins on the square sub-lattice are rotated by the
angle~$q$, the incommensurate ordering vector yields the ratio $J_1 / J_2  =
-2 \cos(q) = -2 \cos(\vec Q \cdot \vec a_{J_2}/2) = 0.47$. Their comparison with
theories including the quantum fluctuations, which are enhanced by the geometric
frustration of the model, suggest even higher values, namely
$J_1/J_2 \approx 0.56$ for large-$N$ Sp($N$) theory~\cite{Chung}, and
$J_1/J_2 \approx 0.72$ for zero-temperature series expansions~\cite{ZhengZeroSeries}.
While the high correction compared to the classical value found in
zero-temperature series expansions is questioned by the authors in their
later study~\cite{Zheng}, the real coupling ratio seems to be above the ratio
$J_1/J_2 = 0.3(2)$ for the optimal fit we found for valley (a) and closer to the
ESR value in figure \ref{fig:residuals}.

% ============================================================================
\section{Extensibility}
% ============================================================================
The presented applications for Heisenberg models are built with a modular
C++ framework for high-temperature series expansions we developed.
The framework provides generic algorithms for the graph generation, the high
temperature expansion and the graph embeddings.
With little effort the presented application could be generalized to include
Dzyaloshinskii-Moriya terms or new applications for fermionic lattice
models, such as the $t$-$J$ can be written based on the existing implementation
of the algorithms.

% ============================================================================
\section{Conclusion}
% ============================================================================

We presented a set of applications and Mathematica packages to compute and
analyze high-temperature series
for the free energy, the uniform magnetic susceptibility and the static
structure factor of Heisenberg models on arbitrary lattices with up to four
independent coupling constants.
As an example we computed the high-temperature series for the anisotropic
triangular lattice and showed how to obtain estimates for the coupling constants
by fitting the Pad\'e approximants to susceptibility data for Cs$_2$CuBr$_4$.

While the obtained high-temperature series for the anisotropic triangular
lattice can be directly applied to other realizations of the same model,
such as Cs$_2$CuCl$_4~$\cite{Coldea2} or various organic superconducting
materials~\cite{McKenzie,Zheng},
the computer programs allow easy access to high-temperature series of
all quantum magnets, in arbitrary dimensions, which can be described by
spin-$1/2$ Heisenberg models on regular lattices, including layered quasi
two-dimensional lattices or dimerized systems.
Therefore the presented applications may prove to be a powerful tool in the lab,
to gain a better understanding experimental results.

% ============================================================================
\section*{Acknowledgments}
% ============================================================================

This project was supported by grants from the Swiss National Supercomputing Centre
(CSCS) under project ID S556, the Paul Scherrer Institute, Deutsche
Forschungsgemeinschaft through the research fellowship WE-5803/1-1 and the ERC
Advanced Grant SIMCOFE.
Computations were performed on Piz Daint at CSCS.

\begin{table*}[t]
\begin{center}
\footnotesize
\begin{minipage}[t]{0.32\textwidth}
\begin{tabular}[t]{cc|c}
$n$ & $m$ & $a_{n,m}$ \\ \hline
0 & 0 & +1\\
0 & 1 & -1/2\\
0 & 3 & +1/24\\
0 & 4 & +5/384\\
0 & 5 & -7/1280\\
0 & 6 & -133/30720\\
0 & 7 & +1/4032\\
0 & 8 & +1269/1146880\\
0 & 9 & +3737/18579456\\
0 & 10 & -339691/1486356480\\
0 & 11 & -1428209/13624934400\\
0 & 12 & +18710029/560568729600\\
1 & 0 & -1\\
1 & 1 & +1\\
1 & 2 & -1/4\\
1 & 3 & -1/12\\
1 & 4 & +1/64\\
1 & 5 & +23/960\\
1 & 6 & +67/46080\\
1 & 7 & -1271/215040\\
1 & 8 & -361/215040\\
1 & 9 & +22433/18579456\\
1 & 10 & +892291/1238630400\\
1 & 11 & -2696291/16349921280\\
2 & 0 & +1/2\\
2 & 1 & -11/16\\
2 & 2 & +25/64\\
2 & 3 & -1/16\\
2 & 4 & -151/3840\\
2 & 5 & +77/9216\\
\end{tabular}
\end{minipage}
\begin{minipage}[t]{0.32\textwidth}
\begin{tabular}[t]{cc|c}
$n$ & $m$ & $a_{n,m}$ \\ \hline
2 & 6 & +29119/2580480\\
2 & 7 & -14033/10321920\\
2 & 8 & -165209/46448640\\
2 & 9 & -30397/530841600\\
2 & 10 & +8128031/7431782400\\
3 & 0 & -1/6\\
3 & 1 & +19/96\\
3 & 2 & -1/6\\
3 & 3 & +1607/11520\\
3 & 4 & -251/6144\\
3 & 5 & -2279/71680\\
3 & 6 & +92311/7741440\\
3 & 7 & +2375329/185794560\\
3 & 8 & -18569821/7431782400\\
3 & 9 & -483818413/98099527680\\
4 & 0 & +13/192\\
4 & 1 & -17/192\\
4 & 2 & +581/15360\\
4 & 3 & -1343/61440\\
4 & 4 & +39833/1032192\\
4 & 5 & -34151/1720320\\
4 & 6 & -654169/61931520\\
4 & 7 & +123366899/14863564800\\
4 & 8 & +3060454129/653996851200\\
5 & 0 & -71/1920\\
5 & 1 & +191/1920\\
5 & 2 & -4319/46080\\
5 & 3 & -7/36864\\
5 & 4 & +204653/5160960\\
5 & 5 & +3095149/309657600\\
\end{tabular}
\end{minipage}
\begin{minipage}[t]{0.32\textwidth}
\begin{tabular}[t]{cc|c}
$n$ & $m$ & $a_{n,m}$ \\ \hline
5 & 6 & -33720647/1238630400\\
5 & 7 & -95689567/18166579200\\
6 & 0 & +367/46080\\
6 & 1 & -219/10240\\
6 & 2 & +1129/26880\\
6 & 3 & -669791/15482880\\
6 & 4 & +11687/2903040\\
6 & 5 & +4785047/206438400\\
6 & 6 & -1201172057/326998425600\\
7 & 0 & +811/322560\\
7 & 1 & -1369/61440\\
7 & 2 & +30497/737280\\
7 & 3 & -292231/37158912\\
7 & 4 & -60070847/1486356480\\
7 & 5 & +3295644749/163499212800\\
8 & 0 & +8213/5160960\\
8 & 1 & -61993/10321920\\
8 & 2 & -23951/123863040\\
8 & 3 & +19321523/928972800\\
8 & 4 & -3477829/185794560\\
9 & 0 & -1729/829440\\
9 & 1 & +317581/23224320\\
9 & 2 & -249945907/7431782400\\
9 & 3 & +863358541/30656102400\\
10 & 0 & -2514101/3715891200\\
10 & 1 & +38886349/7431782400\\
10 & 2 & -50084021/5945425920\\
11 & 0 & +27560147/40874803200\\
11 & 1 & -348043/63866880\\
12 & 0 & +870952109/1961990553600\\
\end{tabular}
\end{minipage}
\caption{Series coefficients of $\chi/ \beta = \frac{1}{4}\sum_{n,m} a_{n,m} (\beta J_1)^n
    (\beta J_2)^m$}
\label{tab:susseries}
\end{center}
\end{table*}

%\section*{References}
\bibliographystyle{elsarticle-num}
\bibliography{refs}{}

\end{document}